\documentstyle[12pt,psfig]{article}
\begin{document}

\newcommand{\gsim}{\mbox{ \raisebox{-1.0ex}{$\stackrel{\textstyle >}
{\textstyle \sim}$ }}}
\newcommand{\lsim}{\mbox{ \raisebox{-1.0ex}{$\stackrel{\textstyle <}
{\textstyle \sim}$ }}}

\begin{flushright}
  \begin{tabular}[t]{l} 
  KEK-TH-593\\
  September 1998
 \end{tabular}
 \end{flushright}
\vspace*{0.5cm}
\centerline{\tenbf  Flavor Physics in the Supergravity Model
\footnote{
    Talk presented
    at International Seminar ``Quarks' 98'', Suzdal, Russia, 
    May 17 - 24, 1998.
}
}
\vspace{0.8cm}
\centerline{\tenrm YASUHIRO OKADA}
\centerline{\tenit KEK}
\centerline{\tenit Oho 1-1, Tsukuba 305, Japan}
\vspace{0.9cm}
Phenomenological aspects of flavor changing processes are 
considered in the context of the supergravity model.
Various flavor changing neutral current processes 
in B and K decays are calculated in such models. 
For lepton flavor violating processes 
the $\mu^+ \rightarrow e \gamma$ branching
ratio and the T odd triple vector correlation for the
$\mu^+ \rightarrow e^+ e^+ e^-$ process are investigated
in the SU(5) SUSY GUT. Possibility to find SUSY effects through
these Flavor changing processes
in future experiments are also discussed.


\section{Introduction}

Present understanding of the elementary particle physics is based on
the gauge theory of quarks and leptons which is called the Standard
Model (SM). Although the SM describes experimental results very well
up to the present available energy scale, it is possible that new
physics appears just above this energy scale. One of the most promising
candidates of physics beyond the SM is unified theory based on 
supersymmetry (SUSY), therefore SUSY particle and SUSY Higgs boson
searches are the most important targets of the present and 
future collider experiments.
    
In order to explore SUSY indirect search experiments are also 
important. For example, through flavor changing neutral current
(FCNC) processes and CP violation in B and K meson decays it may be
possible to identify new physics effects. Also processes like proton
decay, lepton flavor violation (LFV) such as $\mu \rightarrow e
\gamma$ and neutron and electron electric dipole moments (EDM)
are important because these are either forbidden or strongly 
suppressed within the SM.

In this talk we discuss FCNC processes in B and K decays 
such as  $b \rightarrow s \gamma$, $B^0 - \bar{B^0}$ mixing,
the CP violating neutral kaon mixing parameter $\epsilon_K$
and $K \rightarrow \pi \nu \bar{\nu}$ in the supergravity 
model and LFV processes in the SUSY GUT. In the context
of SUSY models flavor physics has important implications. Since 
these processes depend on the structure of the squark and 
slepton mass matrices we may be able to get some insight
on the SUSY breaking mechanism. In fact general SUSY breaking
terms tend to induce too large FCNC and LFV. In the minimal 
supergravity model we assume that the scalar mass terms have 
universal structure at the Planck 
scale and therefore there are no FCNC effects nor LFV  from
the squark and slepton sector at this scale. The physical squark
and slepton masses are determined taking account of the renormalization
effects from the Planck to the weak scale. This will induce sizable
SUSY contributions to various FCNC processes in the B and K decays
\cite{BBMR}.
Also if there is interaction which breaks lepton flavor conservation
between the Planck and the weak scales, the LFV effects can be induced
in the slepton mass matrices. 

In this talk after short introduction to the flavor problem in the
SUSY model we discusses the results of numerical analysis for FCNC 
and LFV processes within the context of the supergravity model.

\section{Flavor Problem in SUSY Models}

As we mentioned in Introduction the squark and the slepton
mass matrices becomes new sources of flavor mixings in the 
SUSY model and generic mass matrices would induce too large FCNC 
and LFV effects if the superpartners' masses are in the 100 GeV
region. If we assume that the SUSY contribution to the  
$K^0 - \bar{K}^0$ mixing is suppressed because of the cancellation
among the squark contributions of different generations, the squarks
with the same $SU(3)\times SU(2)\times U(1)$ quantum numbers have 
to be highly degenerate in masses. When the squark mixing angle is 
in a similar magnitude to the Cabibbo angle the requirement on 
degeneracy becomes as
\begin{equation}
\frac{\Delta m_{\tilde{q}}^2}{ m_{\tilde{q}}^2}\lsim 10^{-2}
\left(\frac{ m_{\tilde{q}}}{100 GeV}\right)
\end{equation}
for at least the first and second generation squarks.
Similarly, the $\mu^+ \rightarrow e^+ \gamma$ process
puts a strong constraint on the flavor off-diagonal terms
for slepton mass matrices which is roughly given by
\begin{equation}
\frac{\Delta m_{\tilde{\mu}\tilde{e}}^2}{ m_{\tilde{l}}^2}\lsim 10^{-3}
\left(\frac{ m_{\tilde{l}}}{100 GeV}\right)^2 .
\end{equation}

In the SUSY model based on the supergravity these flavor problems
can be avoided by setting SUSY breaking mass terms universal at the 
very high energy scale. In fact all the scalar fields are assumed to have 
the same SUSY breaking mass at the Planck scale in the minimal 
supergravity model and therefore there are no FCNC and LFV at this 
scale. Physical squark and slepton masses are, however, defined at the
weak scale and these masses are determined through the renormalization
group equations (RGE). As a result we can derive:\\
\noindent
(1) Squarks for the first and second generations remain highly 
degenerate so that the constraint from the $K^0 - \bar{K}^0$ mixing can 
be safely satisfied.\\  
\noindent
(2) Due to the effect of large top Yukawa coupling constant the stop
and the sbottom can be significantly lighter than other squarks. This
will induce sizable contributions to FCNC processes such as 
$b \rightarrow s\gamma$\cite{BBMR,GO}, $b \rightarrow sl^+l^-$
\cite{BBMR,bsll}, $\Delta M_B$\cite{BBMR,GNO},$\epsilon_K$\cite{GNO} 
and $K \rightarrow \pi \nu \bar{\nu}$.\\
\noindent
(3) In the SUSY GUT the large top Yukawa coupling constant also 
induces the flavor mixing in the slepton sector so that LFV processes
such as $\mu^+ \rightarrow e^+ \gamma$, $\mu^+ \rightarrow e^+e^+e^-$ 
and $\mu^- - e^- $ conversion in atoms receive large SUSY contributions
\cite{BH}.

In the following we discuss the results of numerical analysis for the 
processes listed in (2)\cite{GOS} and (3)\cite{HNOST,OOS}.

\section{B and K decays and the Supergravity Model}

In the minimal SM various FCNC processes and CP violation in B and K decays
are determined by the Cabibbo-Kobayashi-Maskawa (CKM) matrix. The constraints
on the parameters in the CKM matrix elements can be conveniently expressed
in terms of the unitarity triangle. With CP violation at B factory as well as
rare K decay experiments we will be able to check consistency of the 
unitarity triangle and at the same time search for effects of physics 
beyond the SM. In order to distinguish possible new physics effects it 
is important to identify how various models can modify the SM predictions. 
Although general SUSY models can change the lengths and the angles of the 
unitarity triangle  in variety ways, the supergravity model predicts a 
specific pattern of the deviation from the SM. Namely, we can show that 
the SUSY loop contributions to FCNC amplitudes approximately have the 
same dependence on the CKM elements as the SM contributions. 
In particular, the complex phase of the $B^0 - \bar{B^0}$ mixing amplitude 
does not change even if we take into account the SUSY and the charged Higgs 
loop contributions. In terms of the unitarity triangle this means 
that the angle measurements through CP asymmetry in B decays determine 
the CKM matrix elements as in the SM case. On the other hand the length 
of the unitarity triangle determined from $\Delta M_B$ and $\epsilon_K$ 
can be modified.

We have calculated the $\Delta M_B$, $\epsilon_K$ and 
branching ratios of $b \rightarrow s\gamma$, $b \rightarrow sl^+l^-$,
$K_L \rightarrow \pi^0 \nu \bar{\nu}$
and $K^+ \rightarrow \pi^+ \nu \bar{\nu}$ in the SUSY model based
on supergravity\cite{GOS}. In the calculation we have used updated 
results of various SUSY search experiments at LEP2 and Tevatron
as well as the next-to-leading QCD corrections in the calculation
of the $b \rightarrow s\gamma$ branching ratio. In Fig. 1
and Fig. 2 we present  $\Delta M_{B_d}$ and 
Br($K_L \rightarrow \pi^0 \nu \bar{\nu})$ in the present model 
normalized by the same quantities calculated in the SM as the function
of the $b \rightarrow s\gamma$ branching ratio. Note that these ratios 
are essentially independent of the CKM parameters because, as 
mentioned above, the SUSY and the charged Higgs boson loop 
contributions have the same dependence on the CKM parameters. 
Although we only present the results for $\Delta M_{B_d}$ and 
Br($K_L \rightarrow \pi^0 \nu \bar{\nu})$,
$\epsilon_K$ and Br($K^+ \rightarrow \pi^+ \nu \bar{\nu}$)
provide the same constraints on the SUSY parameters respectively
because these quantities are almost equal if normalized by the 
SM prediction. We have calculated the SUSY particle spectrum based on
two different assumptions on the initial conditions of RGE.  
The minimal case corresponds to the minimal supergravity where 
all scalar fields have a common SUSY breaking mass at the GUT scale. 
In the second case shown as ``all'' in the figures we enlarge the SUSY 
parameter space by relaxing the initial conditions for the SUSY breaking 
parameters, namely all squarks and sleptons have a 
common SUSY breaking mass whereas an independent
SUSY breaking parameter is assigned for Higgs fields. 

From Fig. 1 and Fig. 2 we can conclude that the $\Delta M_{B_d}$ 
(and $\epsilon_K$) can be enhanced by up to 40\% and Br($K_L \rightarrow 
\pi^0 \nu \bar{\nu})$ (and Br($K_+ \rightarrow \pi^+ \nu \bar{\nu}$))
is suppressed by up to 10\% for extended parameter space and the corresponding 
numbers for the minimal case are 20\% and 3\%. The ratio of two
Higgs vacuum expectation value, $\tan{\beta}$, is 2 for these figures and 
the deviation from the SM turns out to be smaller for large value of    
$\tan{\beta}$.
\begin{figure}
\begin{center}
\mbox{\psfig{figure=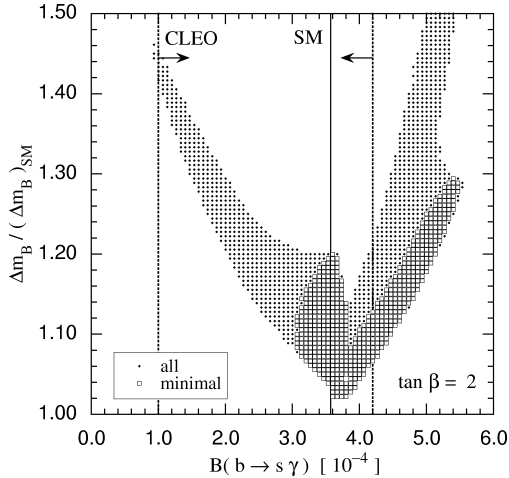,width=2.6in,angle=0}}
\end{center}
\caption{$\Delta M_{B_d}$ normalized by the SM value for $\tan{\beta}=2$
as a function of $b \rightarrow s\gamma$ branching ratio. The square
(dot) points correspond to the minimal (enlarged) parameter space of
the supergravity model. The vertical lines correspond to the CLEO 95 
upper and lower bounds\protect{\cite{CLEO}}.
\label{fig:fig1}}
\end{figure} 
\begin{figure} 
\begin{center}
\mbox{\psfig{figure= 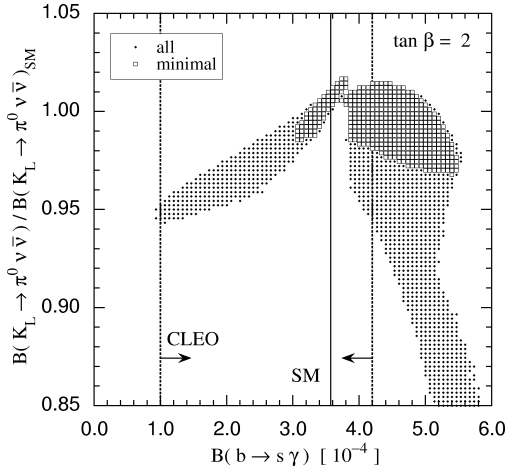,width=2.6in,angle=0}}
\end{center}
\caption{ Br($K_L \rightarrow 
\pi^0 \nu \bar{\nu})$ normalized by the SM value for $\tan{\beta}=2$
as a function of $b \rightarrow s\gamma$ branching ratio.
\label{fig:fig2}}
\end{figure} 

These deviations may be evident in future when B factory experiments 
provide additional information on the CKM parameters. It is expected 
that the one of the three angles of the unitarity triangle is determined 
well through the $B \rightarrow J/\psi K_S$ mode. Then assuming the SM,
one more physical observable can determine the CKM parameters or 
$(\rho,\eta)$ in the Wolfenstein parametrization.
New physics effects
may appear as inconsistency in the determination of these parameters 
from different inputs. For example, the $\rho$ and $\eta$ parameters 
determined from CP asymmetry of B decay in other modes, $\frac
{\Delta M_{B_s}}{\Delta M_{B_d}}$ and $|V_{ub}|$ can be considerably
different from those determined through  $\Delta M_{B_d}$ 
$\epsilon_K$ and Br($K\rightarrow \pi \nu \bar{\nu})$ because
$\Delta M_{B_d}$ $\epsilon_K$ are enhanced and 
Br($K\rightarrow \pi \nu \bar{\nu})$'s are suppressed in the present
model. The pattern of these deviations from the SM will be a key 
to distinguish various new physics effects. We also note from Fig. 1
and Fig. 2 that, although the new results reported at ICHEP98
($2.0\times 10^{-4}<$Br$(b\rightarrow s\gamma)4.5\times 10^{-4}$)
\cite{CLEO98} does not change the situation very much,
future improvement on the $b\rightarrow s\gamma$ branching will give
great impacts on constraining the size of possible deviation from 
the SM in FCNC processes. 

\section{LFV in the SU(5) SUSY GUT }
Another interesting possibility to search for SUSY effects through 
flavor physics is to look for LFV process such as
$\mu^+ \rightarrow e^+ \gamma$, $\mu^+ \rightarrow e^+ e^+ e^-$
and $\mu^-  - e^-$ conversion in atoms. The experimental upper bound on 
these processes quoted in PDG 98 are Br($\mu^+ \rightarrow e^+ \gamma$) 
$\leq 4.9 \times 10^{-11}$, Br($\mu^+ \rightarrow e^+ e^+ e^-$)
$\leq1.0 \times 10^{-12}$ and 
$\frac{\sigma(\mu^-T_i \rightarrow e^-T_i)}
{\sigma(\mu^-T_i \rightarrow caputure)}\leq 4.3 \times 10^{-12}$.
Recently there are considerable interests on these processes because
predicted branching ratios turn out to be close to the upper bounds in 
the SUSY GUT\cite{BH}.  

As discussed in Section 2 no LFV is generated at the Planck scale
in the context of supergravity model. In the SUSY GUT scenario,
however, the LFV can be induced through renormalization effects on 
slepton mass matrix because the GUT interaction breaks lepton flavor
conservation. In the minimal SUSY SU(5) GUT, the effect of the large 
top Yukawa coupling constant results in the LFV in the right-handed 
slepton sector. The numerical calculation shows that there is unfortunate 
cancellation between different diagrams so that  
Br($\mu^+ \rightarrow e^+ \gamma$) is below $10^{-13}$ level
for most of the parameter space\cite{HMTY2}. This is in contrast 
with the SO(10) model where both left- and right-handed sleptons 
induce LFV and the predicted branching ratio is at least larger by two
order of magnitudes\cite{BHS}. 

We have calculated the Br($\mu^+ \rightarrow e^+ \gamma$)
in the context of the SUSY SU(5) model and pointed out that
the branching ratio can be enhanced for large value of
$\tan{\beta}$ once we take into account effects of higher dimensional 
operators to explain realistic fermion masses\cite{HNOST}. 
In the minimal case the Yukawa coupling is given by the superpotential   
$W= (y_u)_{ij}{\bf T_i}\cdot {\bf T_j}\cdot {\bf H(5)}
+(y_d)_{ij}{\bf T_i}\cdot{\bf \bar{F}_j}\cdot {\bf \bar{H}(5)}$
where ${\bf T_i}$ is 10 dimensional and ${\bf \bar{F}_j}$
is 5 dimensional representation of SU(5). It is well known that
this superpotential alone cannot explain the lepton and quark 
mass ratios for the first and second generations although
the $m_b/m_{\tau}$ ratio is in reasonable agreement.
One way to obtain realistic mass ratios are to introduce
higher dimensional operators such as $\frac{f_{ij}}{M_{Planck}}
{\bf \Sigma(24)}\cdot{\bf T_i}\cdot{\bf \bar{F}_j}
\cdot{\bf \bar{H}(5)}$. We investigated how
inclusion of these terms changes prediction of the branching ratio.  
It turns out that the branching ratio is quite sensitive to the
details of these higher dimensional operators. Firstly, once we
include these terms the slepton mixing matrix elements $\lambda_
{\tau}\equiv
V_{\tilde{e}31}^*V_{\tilde{e}32}$ which appear
in the formula of the $\mu^+ \rightarrow e^+ \gamma$
amplitude is no longer related to the corresponding CKM matrix 
elements. More importantly, for large value of $\tan{\beta}$,
the left-handed slepton also induces the LFV and the predicted branching 
ratio becomes enhanced by two order of magnitudes as in the SO(10) 
case\cite{ACH}. The destructive interference among the different diagrams
also disappear. We show one example of such calculation in Fig.3
where Br($\mu^+ \rightarrow e^+ \gamma$) can be close to $10^{-11}$
level for large values of $\tan{\beta}$ in the non-minimal case.
\begin{figure} 
\begin{center}
\mbox{\psfig{figure= 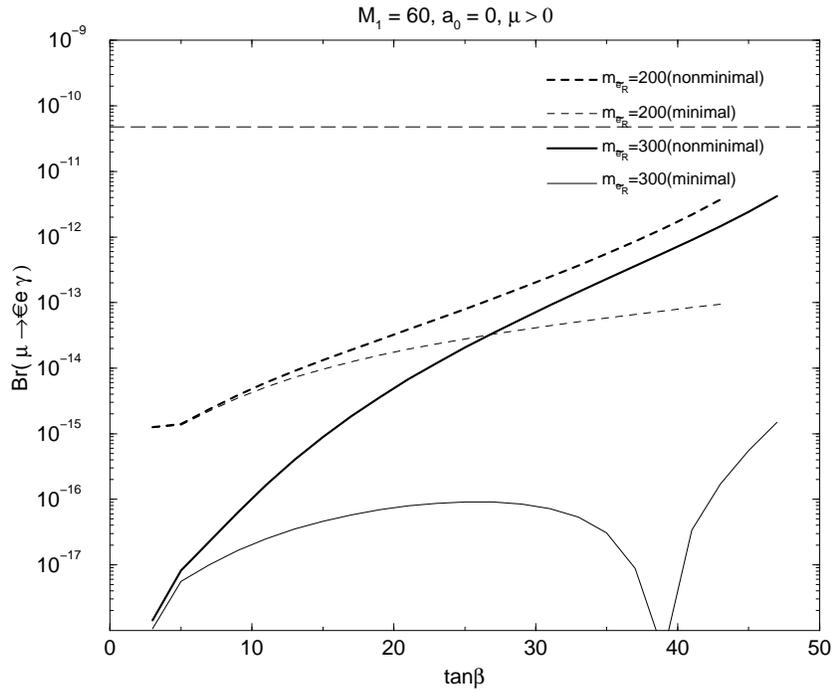,width=5in,angle=-90}}
\end{center}
\caption{
Dependence of the branching ratio of $\mu\rightarrow e \gamma$
on $\tan\beta$ for  the right-handed selectron mass 200GeV 
(dashed lines) and 300GeV (solid lines). The thick lines are for 
the non-minimal case that $V_{\bar{e}}$ and $V_{l}$ are the same as 
$V_{\rm KM}$, and the thin lines are for the minimal case in which 
$V_{\bar{e}}=V_{\rm KM}$ and $V_{l}={\bf 1}$. In this figure we 
choose the bino mass 60GeV, $a_0$ =0, the higgsino mass positive. 
The long-dashed line is the experimental upper bound.
\label{fig:fig3}}
\end{figure} 

Finally, we consider T violating asymmetry in the
$\mu^+ \rightarrow e^+ e^+ e^-$ decay. 
In the polarized muon decay we can define T odd triple vector 
correlation $<\vec{\sigma}\cdot(\vec{p_1}\times\vec{p_2})>$ where
$\vec{\sigma}$ is muon polarization and  $\vec{p_1}$ and
$\vec{p_1}$ are two independent momenta of decay particles\cite{m3eCP}. 
We have investigated possibility of sizable T odd asymmetry in the 
SU(5) SUSY GUT\cite{OOS}. In order to have this asymmetry we need 
to introduce a CP violating phase other than the KM phase. 
In this model the phase can be provided by the complex phases 
in the SUSY breaking terms, for example, the phase
in the triple scalar coupling constant (A term). Since this phase
also induces electron and neutron EDMs, we have calculates the
T odd asymmetry in the $\mu^+ \rightarrow e^+ e^+ e^-$ taking
into account EDM constraints. By numerical calculation we show
that the asymmetry up to 20\% is possible. The branching ratio for 
$\mu^+ \rightarrow e^+ e^+ e^-$ turned out to crucially depend on
the slepton mixing element $\lambda_{\tau}$ which is an 
unknown parameter once we take into account the higher dimensional
operators for the Yukawa coupling constants. 
For $\lambda_{\tau}=10^{-2}$ we can show that 
the branching ratio of $10^{-14}$ is possible with 10\%
asymmetry which can be reached in future experiment with sensitivity
of $10^{-16}$ level.

\section{Conclusions}
We have considered various flavor changing processes in the 
supersymmetric standard model based on the supergravity.
Flavor changing neutral current processes in B and K decays
such as $B^0 - \bar{B^0}$ mixing, $\epsilon_K$ and branching 
ratio of $K \rightarrow \pi \nu \bar{\nu}$ are calculated and 
it is shown that the deviation from the SM becomes as large as 40 \%
for $B^0 - \bar{B^0}$ mixing and $\epsilon_K$ but somewhat smaller
for $K \rightarrow \pi \nu \bar{\nu}$ processes.
We also investigated the lepton flavor violation in the SU(5) SUSY GUT. 
It is pointed out that the $\mu \rightarrow e
\gamma$ branching ratio can be enhanced for large $\tan{\beta}$
if we take into account the higher dimensional operators
in the Yukawa coupling constants at the GUT scale.
The T odd triple vector correlation is also calculated for the
$\mu^+ \rightarrow e^+ e^+ e^-$ process and it is shown that
the asymmetry up to 20\% is possible due to the CP violating
phases in the supersymmetry breaking terms. 
Experiments on B, K and LFV
processes in near future, therefore, will provide very important
opportunities to investigate into the structure of the SUSY 
breaking sector.\\

This work was supported in part by the
Grant-in-Aid of the Ministry of Education, Science, Sports and
Culture, Government of Japan.


%



\newcommand{\Journal}[4]{{#1} {\bf #2}, {(#3)} {#4}}
\newcommand{\pl}{\sl Phys.~Lett.}
\newcommand{\plb}{\sl Phys.~Lett.~{\bf B}}
\newcommand{\prp}{\sl Phys.~Rep.}
\newcommand{\pr}{\sl Phys.~Rev.}
\newcommand{\prd}{\sl Phys.~Rev.~{\bf D}}
\newcommand{\prl}{\sl Phys.~Rev.~Lett.}
\newcommand{\np}{\sl Nucl.~Phys.}
\newcommand{\npb}{\sl Nucl.~Phys.~{\bf B}}
\newcommand{\ptp}{\sl Prog.~Theor.~Phys.}
\newcommand{\zp}{\sl Z.~Phys.}
\newcommand{\zpc}{\sl Z.~Phys.~{\bf C}}
\newcommand{\mpl}{\sl Mod.~Phys.~Lett.}
\newcommand{\rmp}{\sl Rev.~Mod.~Phys.}
\newcommand{\mpla}{\sl Mod.~Phys.~Lett.~{\bf A}}
\newcommand{\sjnp}{\sl Sov.~J.~Nucl.~Phys.}
\newcommand{\ibid}{\it ibid.}



\begin{thebibliography}{99}
\bibitem{BBMR}
  S.~Bertolini, F.~Borzumati, A.~Masiero and G.~Ridolfi,
  \Journal{\np}{B353}{1991}{591}.
\bibitem{GO}
  T.~Goto and Y.~Okada, \Journal{\ptp}{94}{1995}{407} and references
therein.
\bibitem{bsll}
  A.~Ali, G.~Giudice and T.~Mannel, \Journal{\zpc}{67}{1995}{417};
P.~Cho, M.~Misiak and D.~Wyler, \Journal{\prd}{54}{1996}{3329};
T.~Goto, Y.~Okada, Y.~Shimizu and M.~Tanaka,
   \Journal{\prd}{55}{1997}{4273};
J.~Hewett, J.D.~Wells, \Journal{\prd}{55}{1997}{5549}.
\bibitem{GNO}
  T.~Goto, T.~Nihei and Y.~Okada, \Journal{\prd}{53}{1996}{5233};
{\it Erratum}, \Journal{\ibid}{D54}{1996}{5904}.
\bibitem{BH}
  R.~Barbieri and L.J.~Hall, \Journal{\plb}{338}{1994}{212}.
\bibitem{GOS}
  T.~Goto, Y.~Okada, and Y.~Shimizu, KEK-TH-567, hep-ph/9804294,
to be published in Phys. Rev. D.
\bibitem{HNOST}
  J.~Hisano, D.~Nomura, Y.~Okada, Y.~Shimizu and M.~Tanaka, KEK-TH-575,
hep-ph/9805367 to be published in Phys. Rev. D.
\bibitem{OOS}
  Y.~Okada, K.~Okumura and Y.~Shimizu,
\Journal{\pr}{D58}{1998}{051901}.
\bibitem{CLEO}
  CLEO Collaboration, M.S.~Alam {\it et al.},
  \Journal{\prl}{74}{1995}{2885}.
\bibitem{CLEO98}
   CLEO Collaboration, CLEO CONF 98-17.
\bibitem{HMTY2}
  J.~Hisano, T.~Moroi, K.~Tobe and M.~Yamaguchi,
  \Journal{\plb}{391}{1997}{341};
  {\it Erratum}, \Journal{\ibid}{B\ 397}{1997}{357}.
\bibitem{BHS}
        R.~Barbieri, L.~Hall, and A.~Strumia,
        \Journal{\np}{B445}{1995}{219}.
\bibitem{ACH}
        N.~Arkani-Hamed, H.~Cheng, and L.~Hall,
        \Journal{\pr}{D53}{1996}{413}.
\bibitem{m3eCP}
  S.B.~Treiman, F.~Wilczek and A.~Zee, \Journal{\prd}{16}{1977}{152}.
\end{thebibliography}
\end{document}